\documentclass[9pt,journal,compsoc]{IEEEtran} 
\usepackage{amsmath}
\usepackage{amssymb}
\usepackage{amsfonts}
\usepackage{arydshln} 
\usepackage{multirow}
\usepackage{xcolor}
\usepackage{graphicx}
\usepackage{caption}
\usepackage{subfig}
\usepackage{soul}
\usepackage{algorithm}
\usepackage{algorithmicx}
\usepackage[noend]{algpseudocode}

\usepackage[marginal]{footmisc}

\def\BibTeX{{\rm B\kern-.05em{\sc i\kern-.025em b}\kern-.08em
    T\kern-.1667em\lower.7ex\hbox{E}\kern-.125emX}}

\begin{document}

\title{PIMMiner: A High-performance PIM Architecture-aware Graph Mining Framework}


\author{
Jiya Su, 
Peng Jiang, 
Rujia Wang
\IEEEcompsocitemizethanks{
\IEEEcompsocthanksitem J. Su and R. Wang are with the Computer Science Department, Illinois Institute of Technology, Chicago, IL. \protect
E-mail: jsu18@hawk.iit.edu, rwang67@iit.edu.
\IEEEcompsocthanksitem  P. Jiang is with the Computer Science Department, University of Iowa. 
\protect
E-mail: peng-jiang@uiowa.edu.
}
\thanks{}}

\IEEEtitleabstractindextext{
\begin{abstract}
Graph mining applications, such as subgraph pattern matching and mining, are widely used in real-world domains such as bioinformatics, social network analysis, and computer vision. 
Such applications are considered a new class of data-intensive applications that generate massive irregular computation workloads and memory accesses, which degrade the performance significantly.
Leveraging emerging hardware, such as process-in-memory (PIM) technology, could potentially accelerate such applications.
In this paper, we propose PIMMiner, a high-performance PIM architecture graph mining framework. We first identify that current PIM architecture cannot be fully utilized by graph mining applications. Next, we propose a set of optimizations and interfaces that enhance the locality, and internal bandwidth utilization and reduce remote bank accesses and load imbalance through cohesive algorithm and architecture co-designs. 
We compare PIMMiner with several state-of-the-art graph mining frameworks and show that PIMMiner is able to outperform all of them significantly. 
\end{abstract}

}


\maketitle

\IEEEdisplaynontitleabstractindextext
\IEEEpeerreviewmaketitle

\section{Introduction} \label{s:intro}
\textit{Graph mining} (GPMI) is an emerging data mining application that can identify or count subgraphs in graph-structured data based on user-defined patterns. 
GPMI has many real-world use cases, such as motif extraction from gene networks \cite{parthasarathy2010survey}, pattern search over semantic data \cite{elbassuoni2011keyword}, drug discovery in bioinformatics \cite{takigawa2013graph,thafar2020dtigems} and social network analysis \cite{tang2010graph,rao2014new}.
It has attracted extensive attention for performance optimization from systems \cite{wang2018rstream,zaki2015arabesque,shi2020graphpi,jamshidi2020peregrine,mawhirter2021graphzero}, algorithms \cite{mawhirter2019automine,wang2018rstream,zaki2015arabesque} and architecture \cite{besta2021sisa,yao2020locality,rao2020intersectx,chen2022fingers,chen2021flexminer} domain. 
State-of-the-art GPMI algorithms are based on pattern enumeration \cite{mawhirter2019automine,shi2020graphpi,mawhirter2021graphzero}, which search from each vertex in the input graph and perform set operations on the neighbor list to extend the next candidate vertex into the subgraph. The pattern-aware GPMI algorithm shows significant speedup over prior approaches\cite{wang2018rstream,zaki2015arabesque}, showing the great potential to be further accelerated with hardware\cite{besta2021sisa,chen2021flexminer,rao2020intersectx,chen2022fingers}.

Unlike traditional graph processing algorithms, GPMI is more challenging in several ways. First, the computation involves more complex iterations, which may cause severe load imbalance\cite{su2021exploring}. The iteration length depends on graph data (e.g., vertex degree), which is not known until runtime, making the workload scheduling challenging.  
Second, the computation involves enormous data access. The GPMI algorithms need to frequently access neighbor lists of vertices, causing irregular memory accesses\cite{rao2020intersectx,besta2021sisa,talati2022ndminer,dai2022dimmining}. 
Therefore, accelerating GPMI with conventional hardware such as CPU cannot achieve good performance.

Process-in-memory (PIM)~\cite{ahn2015pim} is considered a promising solution to enhance the performance of memory-bounded data-intensive applications. With PIM architecture, it is possible to integrate general-purpose or specialized computation units in or near the memory module. When the application and data are appropriately placed and scheduled on the PIM and host CPU, we can reduce massive data movement between the CPU and memory module to achieve high-performance and energy-efficient computation.
For example, classical graph processing applications, such as BFS and page rank, have been implemented on emerging PIM architectures with software and hardware co-designs~\cite{ahn2015pim,nai2017graphpim}. 
SISA is the first work \cite{besta2021sisa} that uses specialized instruction and PIM hardware to accelerate GPMI applications. 
While the performance gain of SISA is significant, we find that SISA focuses on the optimization of set operation computation while ignoring the memory characteristics and workload distribution on PIM.
Recently, NDMiner \cite{talati2022ndminer}, and DIMMining \cite{dai2022dimmining} were proposed to accelerate GPMI on PIM or near data processor (NDP); however, they also focus on specialized ISA and accelerator design to accelerate the set-centric operations.


We believe that PIM architecture has great potential to accelerate GPMI even with general-purpose ISAs. The preliminary characterization results show that the current PIM architecture is not well-utilized and could benefit from application-aware designs.
In this work, we propose \textit{PIMMiner}, a high-performance PIM architecture-aware graph mining framework. 
We analyze and identify that GPMI applications running on PIM architecture show high load imbalance and cannot fully gain benefits from near-memory accesses.
We propose a set of lightweight, general, and effective optimizations and interfaces in the PIMMiner framework.
First, we incorporate simple filter logic in PIM unit to execute conditions and filter out unnecessary data access in the program, reducing remote memory access significantly.
Next, we propose a locality-enhancing PIM-friendly memory address mapping scheme that allows PIM cores to access near-core memory banks more frequently.
Third, we design an efficient PIM-side workload stealing scheduler with general and GPMI-specific programming interfaces, which can work seamlessly with distributed PIM cores without the shared cache.
Lastly, we discussed the selective data duplication optimization that can be incorporated with our GPMI interface, which can further reduce the remote memory accesses.
By putting all pieces together, our proposed PIMMiner can outperform all existing software and hardware GPMI systems.
On average, our PIMMiner achieves 549x speedup over GraphPi\cite{shi2020graphpi}, and 710x speedup over AutoMine\cite{mawhirter2019automine}. PIMMiner also achieves high speedup over other hardware accelerators and PIM frameworks: 59x speedup over NDMiner\cite{talati2022ndminer} and 2.7x speedup over DIMMining\cite{dai2022dimmining}.

\section{Background} \label{s:background}
\subsection{Preliminaries on Graph Mining}
\label{subsec:GPMI}

\noindent \textit{Graph basics.}
A \emph{graph} $G(V,E)$ includes a set of vertices $V$ and a set of edges $E$.
A graph $G'(V', E')$ is a \emph{subgraph} of graph $G(V, E)$ if $V'\subseteq V$, $E'\subseteq E$. 
Two graphs $G_a=(V_a, E_a)$ and $G_b=(V_b, E_b)$ are isomorphic if there is a bijective function $f: V_a\Rightarrow V_b$ such that $(v_i, v_j)\in E_a$ if and only if $(f(v_i), f(v_j))\in E_b$.
We say two (sub)graphs have the same \emph{pattern} if they are isomorphic. The pattern is a template for the isomorphic subgraphs, and a subgraph is an instance (also called \emph{embedding}) of its pattern. 

\subsubsection{GPMI applications.}
\begin{figure}[t]
\centering 
\includegraphics[width=1\linewidth]{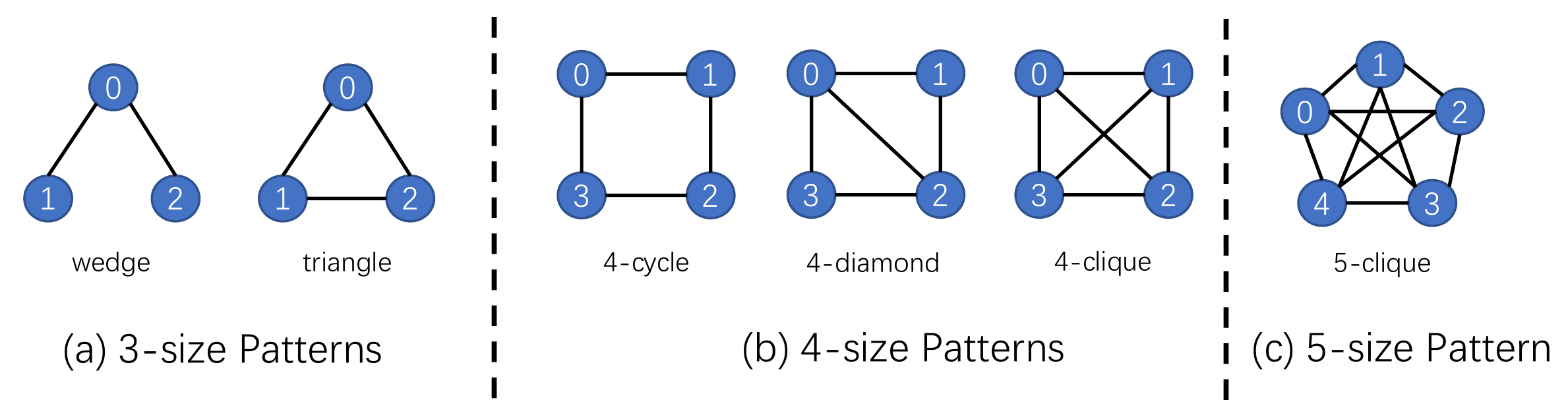}
\vspace{-0.4cm}
\caption{Representative GPMI patterns.} 
\label{fig:pattern}
\vspace{-0.3cm}
\end{figure}

Graph pattern mining needs to find all subgraphs with different patterns that meet the application requirements.
\textit{Motif counting} (MC) is one of the most representative applications of GPMI.
A motif is any connected, unlabeled graph pattern. The goal of motif counting is to identify all motifs (patterns) with \textit{k} vertices and count the embeddings of each pattern. This kernel is widely used in bioinformatics.
There are two patterns with 3 vertices for MC, which are shown in Figure~\ref{fig:pattern}(a). 
\textit{Clique Counting} (CC) counts all the embeddings of the \textit{k}-clique pattern in an unlabeled graph. 
\textit{k}-clique pattern is defined as a fully connected pattern with \textit{k} vertices where each vertex is connected to all other vertices. 
Figure~\ref{fig:pattern} shows size 3 to 5 clique patterns.
There are three 4-size patterns, as shown in Figure~\ref{fig:pattern}(b), including 4-CL(cycle), 4-DI(diamond), and 4-CC(clique).

\subsubsection{Representative GPMI algorithms.} 



\begin{figure}[t]
\centering 
\includegraphics[width=0.9\linewidth]{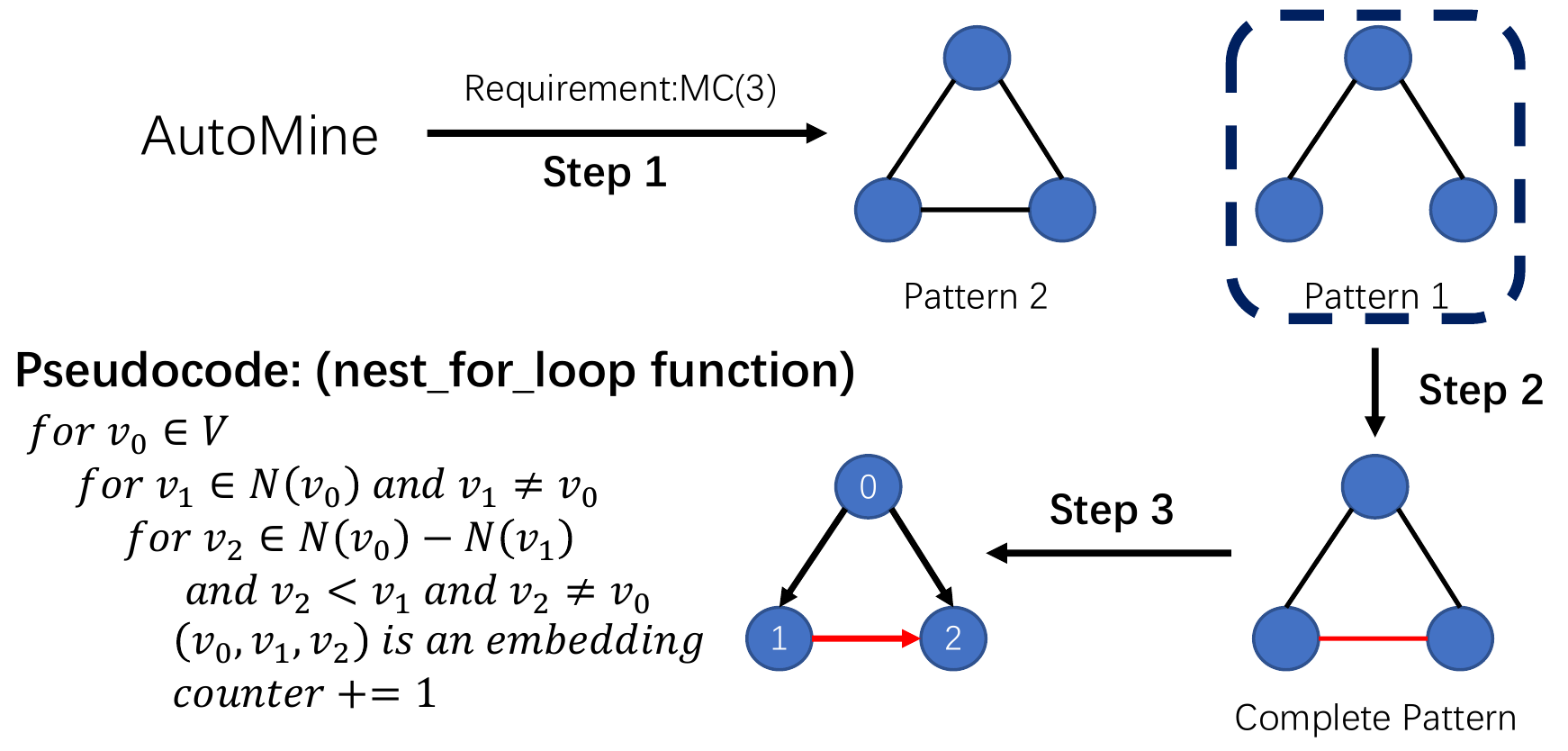} 
\vspace{-0.2cm}
\caption{Pattern enumeration with AutoMine\cite{mawhirter2019automine} method.} 
\label{fig:AutoMine}
\vspace{-0.2cm}
\end{figure}

Pattern-enumeration\cite{mawhirter2019automine,mawhirter2019graphzero,shi2020graphpi} is the state-of-the-art GPMI algorithm since it eliminates the computation-intensive isomorphism tests with lots of edge-dimension random accesses and avoids checking the subgraphs not matching the pattern.
Figure~\ref{fig:AutoMine} shows the steps of pattern enumeration in AutoMine\cite{mawhirter2019automine} algorithm. 

First, it generates all patterns according to the requirements of the application (Step 1).
Then, for each pattern,
it first constructs a colored complete pattern graph to encode all the neighborhood relations of the vertices in the pattern (Step 2). Specifically, it paints all present edges black and adds red edges for the absent ones.
Next, it assigns an order to the vertices of the pattern, and specifies the direction for the edges from small id vertices to large id vertices (Step 3).
Finally, according to the vertex ids and the directed edges, we can construct a multi-layer \textit{nest\_for\_loop} to find all embeddings (also called subgraphs) that match the pattern.
Each vertex in the pattern is associated with a \textit{for} loop. The loops start from the smallest vertex id $v_0$.
If the incoming edge $(i, j)$ is black, which means there is an edge between vertices $i$ and $j$, then vertex $j$ belongs to the \textit{intersection} of the neighbor sets of vertex $i$; if the edge is red, which means there is no edge between vertices $i$ and $j$, then vertex $j$ belongs to the \textit{subtraction} of the neighbor set of the vertex $i$.
Take $v_2$ in the 3-size pattern 1 as an example, since the incoming edge $(0, 2)$ is black and edge $(1, 2)$ is red, $v_2 \in N(v_0) - N(v_1)$.
In addition, to avoid finding duplicate embeddings, according to the matching pattern, the algorithm will add some restrictions to do the symmetry breaking.
For example, in Figure~\ref{fig:AutoMine}, $v_2 < v_1$ is the restriction of pattern 1.

\subsection{Process-in-memory Architecture}
\label{subsec:PIM}

Processing-in-Memory (PIM) integrates processing units inside the memory to reduce the overhead of frequent data movement.
PIM can be implemented using a variety of technologies. 
For example, UPMEM provides a PIM prototype that is DDR4 compatible\cite{devaux2019true}. 
The PIM cores of UPMEM are on the buffer chip of the DIMM. However, due to the constraint of the DDR4 protocol, the single PIM DIMM can only provide limited internal bandwidth (e.g., each at 1GB/s). 
In contrast, due to its large bandwidth and energy advantages, 3D-stacking with TSVs technology is a commonly used technology for PIM. 
The PIM cores could be either implemented on the logic die\cite{gao2015practical,das2018towards,ahn2015pim,huangfu2019medal,nai2017graphpim} or in the DRAM banks\cite{kwon202125,besta2021sisa,lee2021hardware}. 
Samsung has recently started manufacturing HBM-PIM chips~\cite{kwon202125,lee2021hardware}. The HBM-PIM adopts the design that incorporates PIM cores inside memory banks. The host CPU can access the HBM-PIM in either PIM mode or normal mode. The host access the HBM with external links, while the PIM cores access the HBM via internal TSVs. 
We believe that HBM-PIM is currently the most promising PIM hardware with its superior internal and external bandwidth. Therefore, in this paper, we follow the architectural properties of HBM-PIM and assume such hardware as our baseline PIM. Our proposed framework could be applied to other PIM hardware with minimal modifications since the optimizations are general.

\begin{figure}[t]
\centering 
\includegraphics[width=1\linewidth]{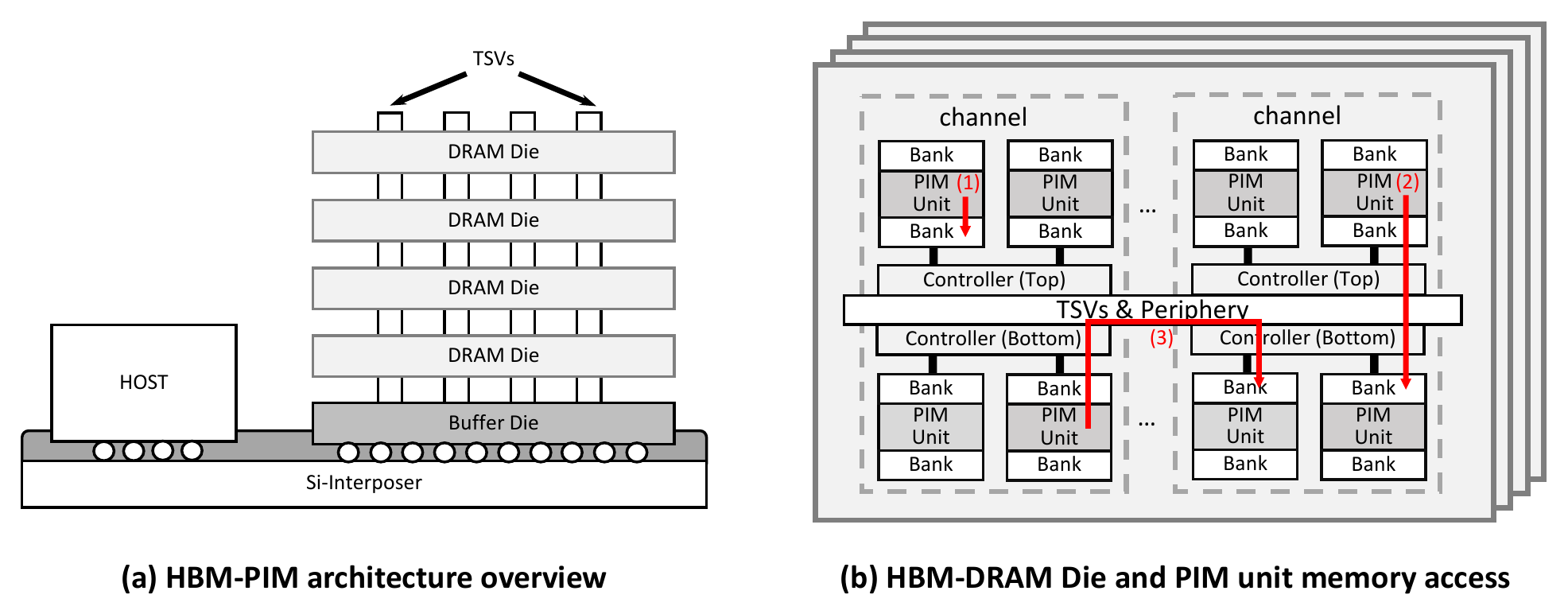}
\vspace{-0.3cm}
\caption{HBM-PIM architecture and internal banks.} 
\label{fig:HBM-PIM}
\vspace{-0.5cm}
\end{figure}

\noindent \textit{HBM-PIM architecture.}
We summarize the key architectural components of HBM-PIM \cite{lee2021hardware} and show the high-level architecture overview in Figure \ref{fig:HBM-PIM}. 
The entire HBM-PIM component can communicate with the host via the silicon interposer.
Each DRAM die can be divided into multiple channels and can be accessed independently.
The PIM execution units are placed at the I/O boundary of a bank. 
As a result, there are three ways for a PIM unit to access memory, as shown in Figure \ref{fig:HBM-PIM}(b): 
(1) near-core bank access, which means that the PIM unit can access its own memory banks with the highest bandwidth and lowest latency; 
(2) intra-channel bank access, which means that the PIM unit can access other memory banks in the same channel via the periphery I/O;
(3) inter-channel remote bank access, which refers to the PIM unit accessing a remote memory bank in other channels via the periphery I/O and TSVs.
The memory access bandwidth and latency are different for the three cases. As a result, how data is placed across banks and channels could significantly impact the PIM performance.


\section{GPMI Characterization on PIM Architecture} \label{s:motivation}
\label{sec:motivation}

In this section, we present and discuss our initial study of using PIM to accelerate GPMI applications. We show the results of mining the 4-CC pattern (other patterns have similar results).




\subsection{Baseline PIM Performance}

We run the 4-CC pattern mining on baseline HBM-PIM architecture and compare the results with CPU in Table~\ref{tab:Performance 128-core CPU and 128-core PIM 4-CC}.
We first sort the graph vertices in the descending order of degree, and then use the round-robin method to assign the tasks to different threads or PIM cores, which is commonly used in related work \cite{talati2022ndminer,su2021exploring}.
In Table~\ref{tab:Performance 128-core CPU and 128-core PIM 4-CC}, our CPU performance is measured on a real system (48 physical cores, 96 hyperthreads, 1152 Gflops), and the PIM performance is measured from a cycle-accurate simulator (128 cores, 1024 Gflops). The details of our experimental methodologies are in Section~\ref{subsec:setup}.

\begin{table}[h]
	\centering  
	\small
	\caption{Performance comparison of a 96-thread CPU and 128-core PIM (4-CC).\protect\footnotemark[1]}  
 
	\label{tab:Performance 128-core CPU and 128-core PIM 4-CC}  
	\resizebox{0.7\linewidth}{!}{
	\begin{tabular}{c|c|c|c}  
	\hline
	Graph & CPU Time (s) & PIM Time (s)  & Speedup\\
	\hline  
	CI & 2.25E-04 & 3.45E-05 & 6.52  \\
        PP & 1.59E-03 & 2.01E-04 & 7.93  \\
        AS & 2.69E-02 & 9.23E-03 & 2.91  \\
        MI & 7.07E-02 & 5.07E-02 & 1.39  \\
        YT & 1.10E-02 & 5.41E-02 & 0.20  \\
        PA & 5.12E-03 & 2.90E-03 & 1.76  \\
        LJ & 1.07E-01 & 1.49E-01 & 0.71  \\
        \hline
	\end{tabular}
	}
	\vspace{-0.1in}
\end{table}
\footnotetext[1]{MI uses 10\% sampling, YT and PA use 1\% sampling, and LJ uses 0.1\% sampling.}

Although the baseline PIM architecture has similar computing throughput, higher memory bandwidth, and lower memory latency compared with the CPU, we find that the performance of PIM is not significantly higher.
For two small graphs \emph{CI} and \emph{PP}, PIM achieves 7x speedup over CPU, but this is mainly due to the fact that these two graphs are small and the overhead of OpenMP dominates the CPU time. 
For other graphs, PIM achieves up to 3x speedup over CPU, which is small compared to the memory advantages on PIM. 
Furthermore, for graphs \emph{YT} and \emph{LJ}, the performance of PIM is even worse than CPU.



\subsection{Memory Access Distribution}
\label{subsec:Memory access distribution on PIM}
To understand the performance bottleneck of PIM architecture, we characterize the proportion of different types of PIM memory accesses in Table \ref{tab:4CC memory access ratio on PIM}.
As described in Section \ref{subsec:PIM}, HBM-PIM has different ways to access memory, and we should optimize the program so that it can access the near-core bank memory as much as possible.
Unfortunately, with the default address mapping of HBM (described in section \ref{subsec:HBM-PIM memory controller}), we find that inter-channel remote memory accesses dominate the memory accesses, which are over 95\% in all cases, which means that the vast majority of our memory accesses require high latency. The memory bandwidth consumed by the program is much lower than the maximum bandwidth provided by the HBM-PIM architecture.


\begin{table}[!ht]
	\centering  
	\small
	\caption{PIM units memory access distribution (4-CC).}  
	\label{tab:4CC memory access ratio on PIM}  
	\resizebox{0.7\linewidth}{!}{
	\begin{tabular}{c|c|c|c}  
	\hline
	Graph & Near-core & Intra-channel  & Inter-channel \\
	\hline  
	CI & 1.29\% & 2.35\% & 96.36\% \\
        PP & 1.41\% & 2.32\% & 96.26\% \\
        AS & 1.70\% & 2.47\% & 95.83\% \\
        MI & 1.30\% & 2.34\% & 96.36\% \\
        YT & 1.43\% & 2.33\% & 96.23\% \\
        PA & 2.05\% & 2.34\% & 95.61\% \\
        LJ & 2.19\% & 2.31\% & 95.50\% \\
        \hline
	\end{tabular}
	}
	\vspace{-0.15in}
\end{table}

\subsection{Load Imbalance on PIM}
\label{subsec:load imbalance}

\begin{figure}[!h]
\vspace{-0.15in}
\centering 
\includegraphics[width=0.8\linewidth]{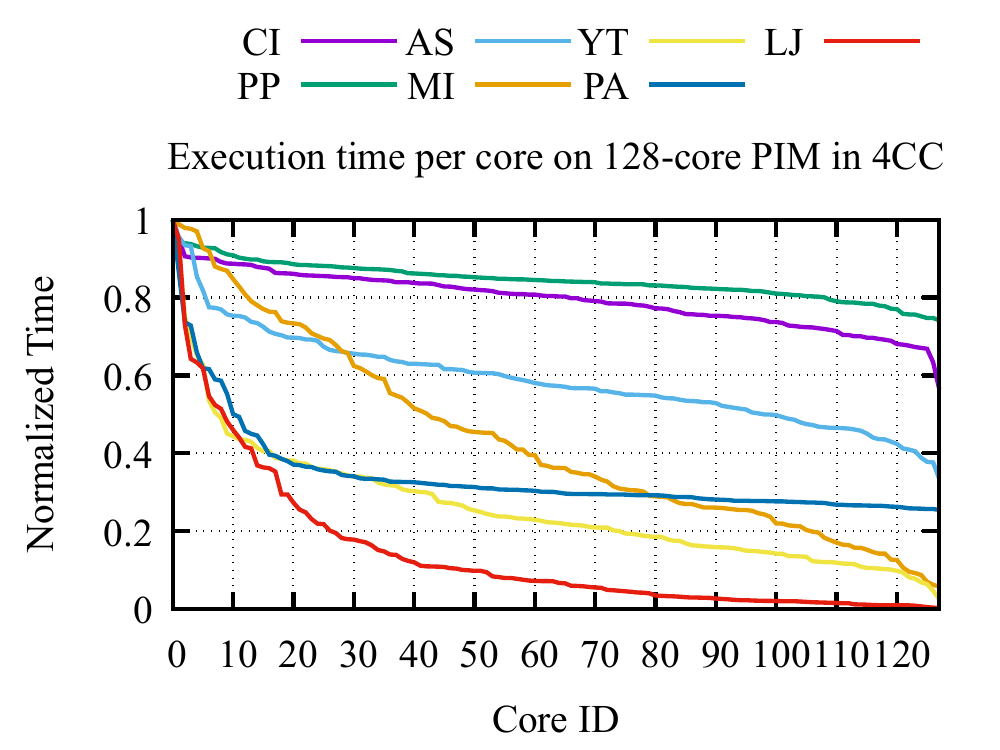} 
 \vspace{-0.1in}
\caption{Load distribution among different cores on PIM (4-CC).} 
\label{fig:imbalance}
 \vspace{-0.1in}
\end{figure}

Additionally, we also characterize the PIM unit utilization ratio by capturing the execution time on PIM cores. We find that GPMI also has severe load imbalance when executing on PIM architecture. 
We plot the load distribution of each PIM core in Figure \ref{fig:imbalance} to show this issue.
For 4-CC, graphs \emph{MI}, \emph{YT}, \emph{PA}, and \emph{LJ} have obvious load imbalance on PIM.

\noindent \textit{Analysis.}
The load imbalance of GPMI mainly comes from the mapping of the loop iterations to the PIM cores. From Figure~\ref{fig:AutoMine}, we know that finding
$n$-size patterns requires $n$ layers of \textit{for} loops. When executing the code on multiple cores, the most straightforward way is to assign the intersection and subtraction (I/S) operations (the second \textit{for} loop to the last \textit{for} loop) to the same core base on the root vertex ($v_0$ in the first-level loop). Such a method (root vertex-based assignment) can guarantee the data dependency of the following I/S operations. 
However, the number of loops at each layer is determined by the \textit{{results}} of I/S operations (e.g., $N(v_0)-N(v_1)$ in Figure~\ref{fig:AutoMine}), which varies a lot based on patterns or graphs, and cannot be determined by offline profiling.
In comparison, for general graph processing applications such as BFS and PR, the workload of each vertex is small and easy to obtain from the vertex degree.
Therefore, compared with general graph processing applications, the workloads of GPMI applications on different cores could differ significantly.
As the matching size increases, the number of layers of the \textit{for} loop increases, resulting in a more significant load imbalance.

\vspace{0.05in}


\noindent \textbf{Key takeaway and motivation.} 
Based on the characterization results, we have the following takeaways.
First, although PIM could theoretically improve GPMI applications with low latency and high bandwidth memory accesses, directly offloading the GPMI execution kernel to PIM cannot guarantee performance speedup. 
Second, for the GPMI application, the inter-channel remote bank access dominates memory access on PIM, which reduces the effectiveness of using PIM to accelerate such applications.
Third, we realize that GPMI has a severe load imbalance issue on PIM units and must be addressed dynamically.
Based on these observations, we are motivated to design a new PIM-aware GPMI framework to improve overall PIM performance by leveraging fast near-core memory accesses and dynamic workload distribution across cores. The framework should support easy-to-use interfaces for programmers and can apply optimizations on demand.

\section{The Design of PIMMiner Framework}
\label{sec:PIMMiner Design}

In this section, we present the PIMMiner framework. Our framework works on top of a system with both HBM and HBM-PIM memory. The two types of HBMs are placed in separate memory channels and can directly interact with the host CPU. 
In addition, the HBM-PIM can directly load data from disk through the DMA, without the need to store data in main memory and then copy the data from main memory to other memory like GPU, thus reducing the communication overhead between main memory and HBM-PIM memory.
The overview of our framework is shown in \S \ref{subsec:overview}. 
Then, we introduce our optimizations, namely application-aware memory access filter (\S \ref{subsec:filter}), memory address management (\S \ref{subsec:HBM-PIM memory controller}), and efficient PIM-size workload stealing scheduler (\S \ref{subsec:steal}).
At last, we describe a set of programming interfaces of PIMMiner to utilize HBM-PIM (\S \ref{subsec:interface}), as well as automatic GPMI interfaces to count the pattern
(\S \ref{subsec:GPMI interface Implementation}).

\subsection{Overview}
\label{subsec:overview}

\begin{figure*}[t]
\centering 
\includegraphics[width=1\textwidth]{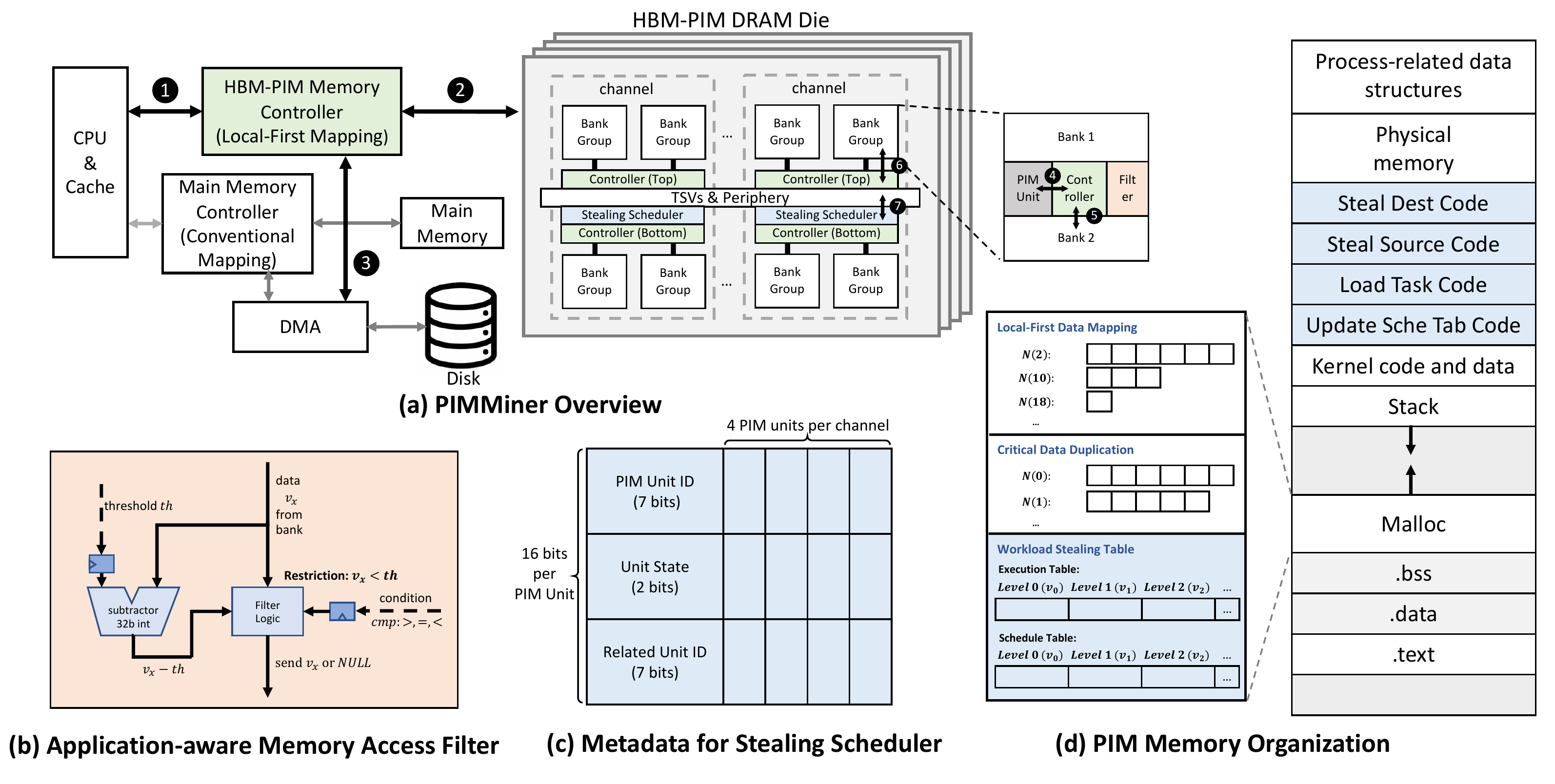}
\vspace{-0.3cm}
\caption{PIMMiner framework.} 
\label{fig:framework}
\vspace{-0.5cm}
\end{figure*}

\subsubsection{System architecture}
We first present the architecture of PIMMiner in Figure \ref{fig:framework}(a). We assume that the HBM-PIM is added to the system in addition to the conventional HBM memory. The memory controllers for the two types of memory are separate. Similar to the conventional HBM access, the HBM-PIM memory controller can support host to HBM-PIM access and direct data transfer from the disk to HBM-PIM through the DMA.

\subsubsection{PIMMiner components}
We highlighted the PIMMiner component with shaded boxes in Figure \ref{fig:framework}(a). The HBM-PIM memory controller supports a dedicated memory address mapping that is beneficial for PIM unit memory accesses. On the HBM-PIM DRAM die, the channel controller includes a new workload stealing scheduler to balance workload across PIM units. Besides, an application-aware memory access filter is incorporated with the PIM unit to reduce unnecessary data access and further reduce data transfers.  


\subsection{Application-aware Memory Access Filter}
\label{subsec:filter}

\noindent \textit{Observation from GPMI applications. }
By analyzing the GPMI algorithm, we find that when accessing the neighbor lists ($N(v)$), the algorithm only needs a part of neighbor vertices for computation due to symmetry breaking.
For example, in Figure~\ref{fig:AutoMine}, the program only needs vertices in $N(v_0)-N(v_1)$ that have smaller id than $v_1$.
In conventional memory architectures, there are no processing units in memory, so unnecessary memory accesses are sent to the host CPU and cannot be removed in advance.
With the PIM architecture, it is possible to let the PIM unit read the data in advance and filter out the data that do not meet the conditions, thereby reducing data transmission.

\noindent \textit{The access filter design. }
The design of the access filter of PIMMiner is very lightweight, as shown in Figure \ref{fig:framework}(b).
One 32-bit filter contains one subtractor, one filter logic, and two registers.
When a bank receives a memory access request to the neighbor list $N(v)$, it also receives a comparison symbol $cmp$ (such as $<$) and the value to be compared ($th$ in Figure \ref{fig:framework}(b)).
Both the comparison symbol $cmp$ and the value $th$ will be stored in two registers (dark blue squares in Figure \ref{fig:framework}(b)) respectively.
Then, the filter reads the data $v_x$ from the sense amplifier.
$v_x$ is first sent to the subtractor and subtracted with $th$ ($v_x-th$).
The result (1 is positive, 0 is equal, and -1 is negative) will be sent to filter logic. In filter logic, if the result matches the comparison symbol $cmp$, output data $v_x$; otherwise, filter out the data. The filter logic can be implemented by a multiplexer. 

\noindent \textit{Timing overhead. }
According to the filtering process, only two cycles of latency are required for each data filter (one cycle for subtraction, and one cycle for comparison).
In our experiment, the width of TSVs on HBM-PIM is 64 bits.
Since one filter can handle a 32-bit integer, to maximize TSV bandwidth utilization, we add two filters per bank group, which can send 64-bit data per clock cycle.

\subsection{Memory Address Management}
\label{subsec:HBM-PIM memory controller}

In Section \ref{subsec:Memory access distribution on PIM}, we find that using the default address mapping will lead to a large number of inter-channel remote memory accesses, which reduces the performance.
PIMMiner modifies the address mapping method in the HBM-PIM memory controller (green box in Figure \ref{fig:framework}(a)), so that the continuous data can be stored in the same bank group, thereby reducing remote bank access.

In this section, we first discuss the problem of default address mapping under the HBM-PIM framework, then propose our improved mapping method \emph{PIM-friendly Local-First Data Mapping} and give the reason why it is suitable for HBM-PIM, and last, we will talk about the impact of local-first data mapping on CPU performance.

\subsubsection{Default address mapping}

Default address mapping is designed for the computing systems where the processors are out of memory, such as CPU and GPU.
To maximize the off-chip memory bandwidth, the default address mapping takes advantage of the parallelism of reading data between different channels, interleaving consecutive memory addresses into different channels, so that when reading a continuous piece of data, the data can be accessed simultaneously in different channels, and then aggregated outside the channels and sent to the processors.
As shown in Figure~\ref{fig:AddressMapping}(a), consecutive memory addresses are first assigned to different channels, then to different banks in the same bank group, and finally to different bank groups.
However, in HBM-PIM, the PIM units are placed near the bank groups.
If the HBM-PIM memory controller uses the default address mapping, when a PIM unit accesses data, it needs to read from different HBM-PIM channels, which greatly increases the inter-channel remote banks accesses, as shown in Table \ref{tab:4CC memory access ratio on PIM}.

\subsubsection{PIM-friendly data mapping}
\begin{figure}[t]
\centering 
\includegraphics[width=\linewidth]{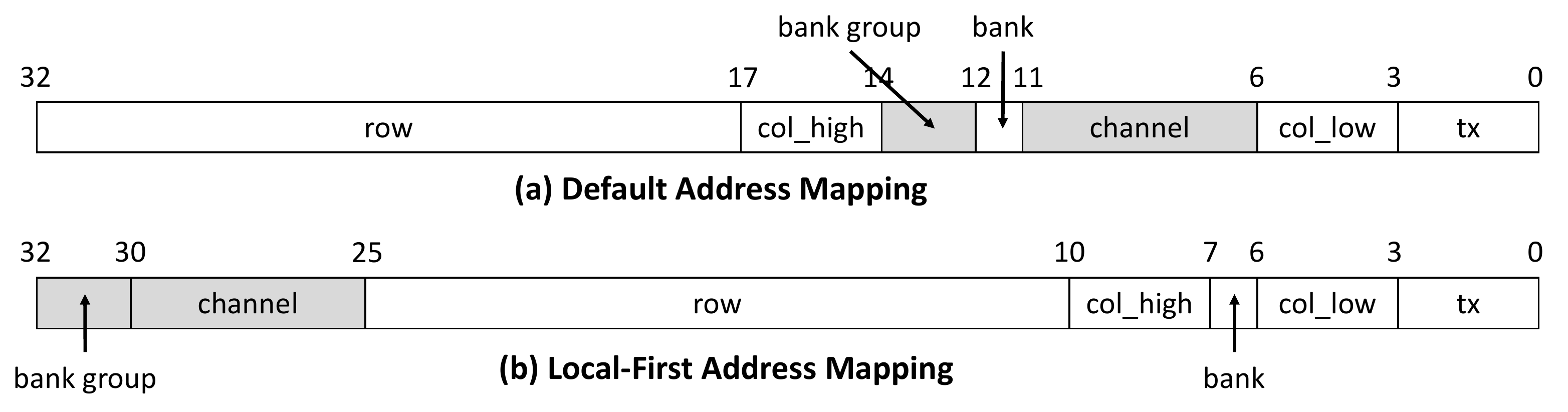}
\vspace{-0.5cm}
\caption{Default and PIM-friendly address mapping.} 
\label{fig:AddressMapping}
\vspace{-0.5cm}
\end{figure}

In PIMMiner, we propose PIM-friendly local-first data mapping, which is implemented in the HBM-PIM memory controller. The consecutive addresses will be first mapped to the banks of the same PIM unit.
As shown in Figure~\ref{fig:AddressMapping}(b), the address first represents the bank group ID, then the channel ID, and finally the different rows, columns, and banks.
To balance the workload between different channels, PIMMiner will first assign PIM unit ID to different channels and then to different bank groups in the same channel.
Because PIMMiner currently uses physical address directly in HBM-PIM memory, based on this mapping method, the compiler can easily allocate data to the bank group of the specified PIM unit, thereby implementing the \emph{PIM\_malloc} interface on CPU in Figure \ref{fig:interface}.

\subsubsection{CPU performance impact}
The local-first data mapping is only implemented at the HBM-PIM memory controller. As a result, normal HBM memory accesses from the CPU will not be affected, as shown in Figure \ref{fig:framework}(a). We believe this design is reasonable because the current PIM prototype \cite{lee2021hardware} does not support concurrent memory accesses from the host CPU and PIM cores.
We also discuss the different access interface designs in Figure \ref{fig:interface}.

\subsection{Efficient PIM-side Workload Stealing Scheduler}
\label{subsec:steal}

Like other multi-core processors, HBM-PIM also has load imbalance issues when processing irregular applications, such as GPMI (\S \ref{subsec:load imbalance}).
Workload stealing is a common method to mitigate load imbalance for general CPU and GPU architectures with its small execution overhead. 
However, directly adopting workload stealing for HBM-PIM architecture is difficult. 
The reason is that,
first, CPUs and GPUs all have shared caches to facilitate work stealing, while HBM-PIM does not. Second, PIM units are scattered in different memory bank groups and channels and cannot directly talk to each other.
As a result, the overhead of stealing a task on HBM-PIM architecture is much higher.
We propose a new workload stealing scheduler in PIMMiner, which can efficiently speed up the stealing process with low overhead.


\subsubsection{Architectural Modifications}

The modifications to support workload stealing on PIM are shown in the blue boxes in Figure \ref{fig:framework} (d).
Each PIM unit maintains two tables during the execution, namely \emph{Execution Table} and \emph{Stealing Table}. 
Among them, \emph{Execution Table} stores the currently executed task ID, and \emph{Schedule Table} stores the task ID to be executed next. 
Additionally, during the preprocessing, the CPU will load \emph{Steal Dest Code}, \emph{Steal Source Code},  \emph{Load Task Code}, and \emph{Update Sche Tab Code} on each PIM unit.
The \emph{Steal Dest Code} and \emph{Steal Source Code}  will update the tables on the destination and source PIM units during the workload stealing. 
\emph{Load Task Code} will load the task from \emph{Schedule Table} to \emph{Execution Table}, and \emph{Update Sche Tab Code} will update the \emph{Schedule Table} to the next task ID.
We will discuss the GPMI implementation of programming interfaces that support workload stealing in section \ref{subsec:GPMI interface Implementation}.

\subsubsection{Stealing Scheduler Design}
The stealing scheduler in PIMMiner is placed on each memory channel, which contains the metadata for workload stealing, as shown in Figure \ref{fig:framework} (c).
For each PIM unit on the same channel, the scheduler will first store the PIM unit ID, and then store the state of the corresponding PIM unit, where 00B means idle, 01B means normal execution, 10B means stealing tasks, 11B means being stolen tasks.
If the PIM unit state is 10B (the PIM unit is stealing tasks), the related unit ID will store the stolen PIM unit ID; if the PIM unit status is 11B (the PIM unit is being stolen tasks), then the related unit ID will store the stealing PIM unit ID.

\subsubsection{Stealing Process}

\begin{figure}[!h]
\centering 
\vspace{-0.3cm}
\includegraphics[width=\linewidth]{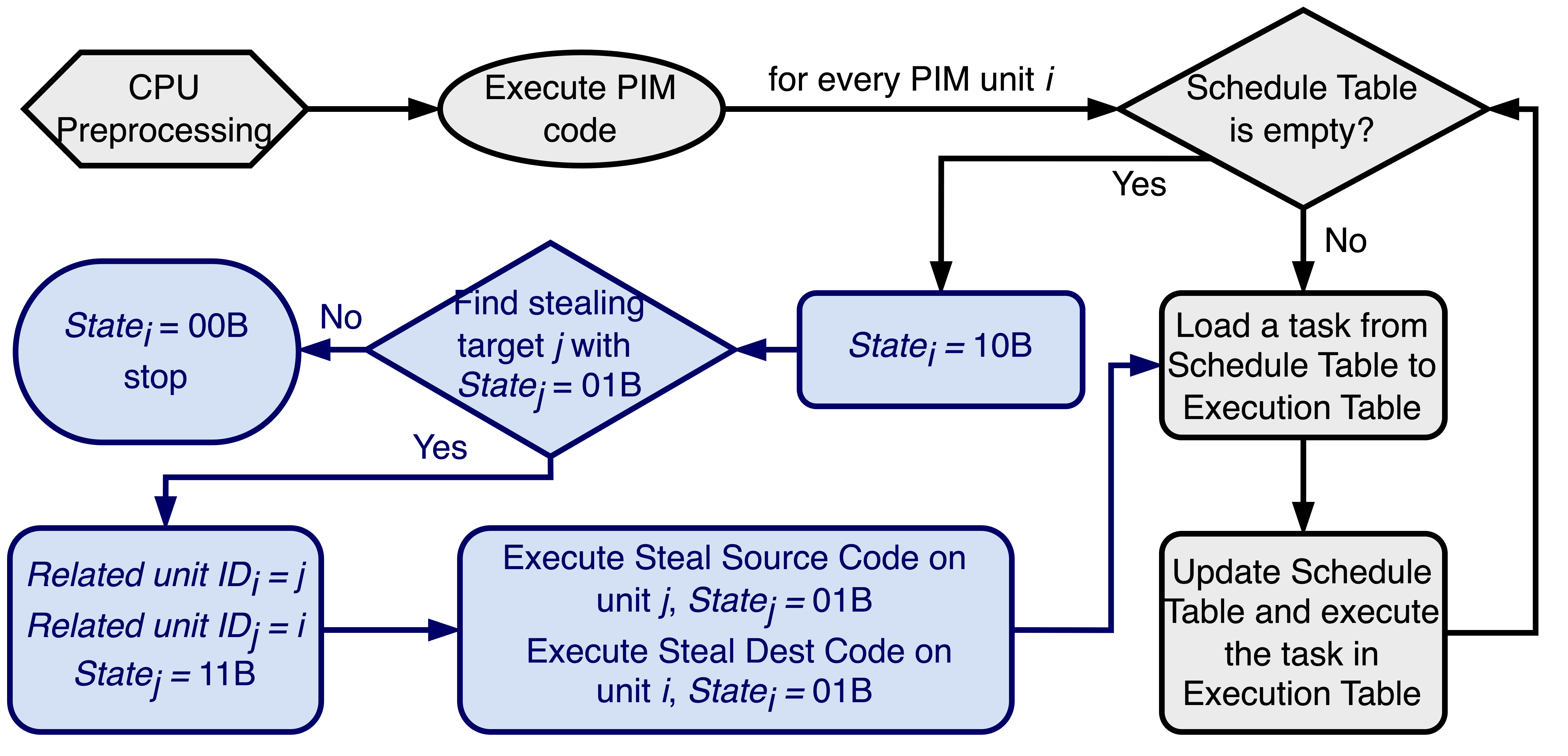}
\vspace{-0.5cm}
\caption{PIMMiner task execution workflow.} 
\label{fig:Stealing workflow}
\vspace{-0.2cm}
\end{figure}

Figure \ref{fig:Stealing workflow} shows the task execution workflow of our PIMMiner. The gray section is the execution logic of the PIM unit and the blue section is the stealing logic.
In the beginning, the PIM unit \emph{i} will check whether \emph{Schedule Table} is empty.
If the \emph{Schedule Table} is not empty, the PIM unit \emph{i} will execute the \emph{Load Task Code} to load the task from \emph{Schedule Table} to the \emph{Execution Table} before executing each task, and then run the \emph{Update Sche Tab Code} to update the schedule table to the next task.
If the \emph{Schedule Table} is empty, which means that PIM unit \emph{i} has finished executing its own tasks, then it will start to steal tasks from other PIM units.

\noindent \textit{Find stealing target: }
During the stealing process, the PIM unit \emph{i} will set the corresponding unit state ($State_{i}$) in the stealing scheduler to 10B, indicating that PIM unit \emph{i} is stealing tasks. 
Then PIM unit \emph{i} will check whether other unit \emph{j} has tasks ($State_j=01B$ ?).
Since each scheduler in PIMMiner only stores the states of PIM units in its own channel, PIM unit \emph{i} will first steal the task from the other PIM unit in the same channel.
If all the unit states in the scheduler are not 01B, that is, all PIM units in this channel have no tasks that can be stolen, then PIM unit \emph{i} will move to the next stealing scheduler, which is in the next channel, and check the unit states in that scheduler.

\noindent \textit{Interactions between stealing source and target units: }
If the stealing target is found, represented as PIM unit \emph{j}, then PIMMiner starts stealing tasks from PIM unit \emph{j} to unit \emph{i}.
PIM unit \emph{i} will first set its own \emph{Related Unit ID} to \emph{j}, indicating that unit \emph{i} will steal the task from unit \emph{j}.
Then, unit \emph{i} will send a stealing signal to unit \emph{j}, and wait for the stealing completion signal from PIM unit \emph{j}. 
After receiving the stealing signal from unit \emph{i}, unit \emph{j} will suspend the current task, and start stealing the tasks from its own \emph{Schedule Table}.
First, unit \emph{j} will set its state to 11B in the scheduler, indicating that unit \emph{j} is stealing its tasks, and other units can no longer send stealing requests to unit \emph{j}.
At the same time, unit \emph{j} will also set its \emph{Related Unit ID} to \emph{i}, which means that it will send the stealing tasks to unit \emph{i} after finishing stealing.
Then, PIM unit \emph{j} will execute \emph{Steal Source Code} to steal tasks from its own schedule table. 
After the code is executed, the stolen tasks will be sent to PIM unit \emph{i} in the form of the schedule table, and then the state will be reset to 01B.
Finally, PIM unit \emph{j} will continue to execute the program that is interrupted before.
After PIM unit \emph{i} receives the schedule table sent by unit \emph{j}, it will execute the \emph{Steal Dest Code} to copy the tasks to its own schedule table, and then set the state in the scheduler to 01B to start executing the tasks.

\noindent \textit{End of the stealing: }
If the unit states in the whole channels are all 10B (the state of stealing tasks), which means that all the units are in the stealing status and PIMMiner has no tasks to execute, then PIM unit \emph{i} will return that the stealing task cannot be found, and the state of unit \emph{i} is set to 00B, indicating that the PIM unit \emph{i} is idle, and finally terminates the running on PIM unit \emph{i}.

\subsubsection{Implementation Details to Support GPMI}
\label{subsubsec:Implementation Details to Support GPMI}
Our workload stealing workflow can be applied to all types of applications. In this section, we describe how to use it to support GPMI applications. We use the GPMI algorithm in Figure \ref{fig:AutoMine} to show how tasks are organized by vertex id from different loop levels.

The \emph{Execution Table} and \emph{Schedule Table}, as shown in Figure \ref{fig:framework} (d), store the index of the neighbor list in each \emph{for loop} to track the tasks.
Among them, \emph{Execution Table} $T_{exe}$ stores the index of the neighbor list in each \emph{for loop}, which is being executed by the PIM unit; and \emph{Schedule Table} $T_{sch}$ stores the index to be executed next, usually $T_{sch}[i] = T_{exe}[i]+1$.
Therefore, when executing the algorithm in Figure \ref{fig:AutoMine},
$v_0 = T_{exe}[0], v_1 = N(v_0)[T_{exe}[1]], v_2 = (N(v_0)-N(v_1))[T_{exe}[2]]$, where $N(v)[i]$ represents the $i$-th neighbor vertex in the neighbor list of vertex $v$.

When loading a task from \emph{Schedule Table}, the PIM unit will first load the last-level index (level 2 in this example) and then move to the lower level if no task is available.
When updating the \emph{Schedule Table}, the PIM unit will also update from the higher level (e.g., level 2) to the lower level (e.g., level 0).

During the stealing, the PIM unit will start from level 0 first in the \emph{Schedule Table}.
If level 0 in the schedule table has a task, that is, the index value is less than the total number of vertices, then the PIM unit will steal the level 0 index from the schedule table to the stealing table, set other levels in the stealing table to 0, and then add one (or the number of PIM units) to the level 0 index in the schedule table.
If level 0 in the schedule table has no tasks, then the PIM unit will steal the level 0 index from the execution table, then steal the index of level 1 in the schedule table, and so on.




\begin{figure}[t]
\centering 
\includegraphics[width=1\linewidth]{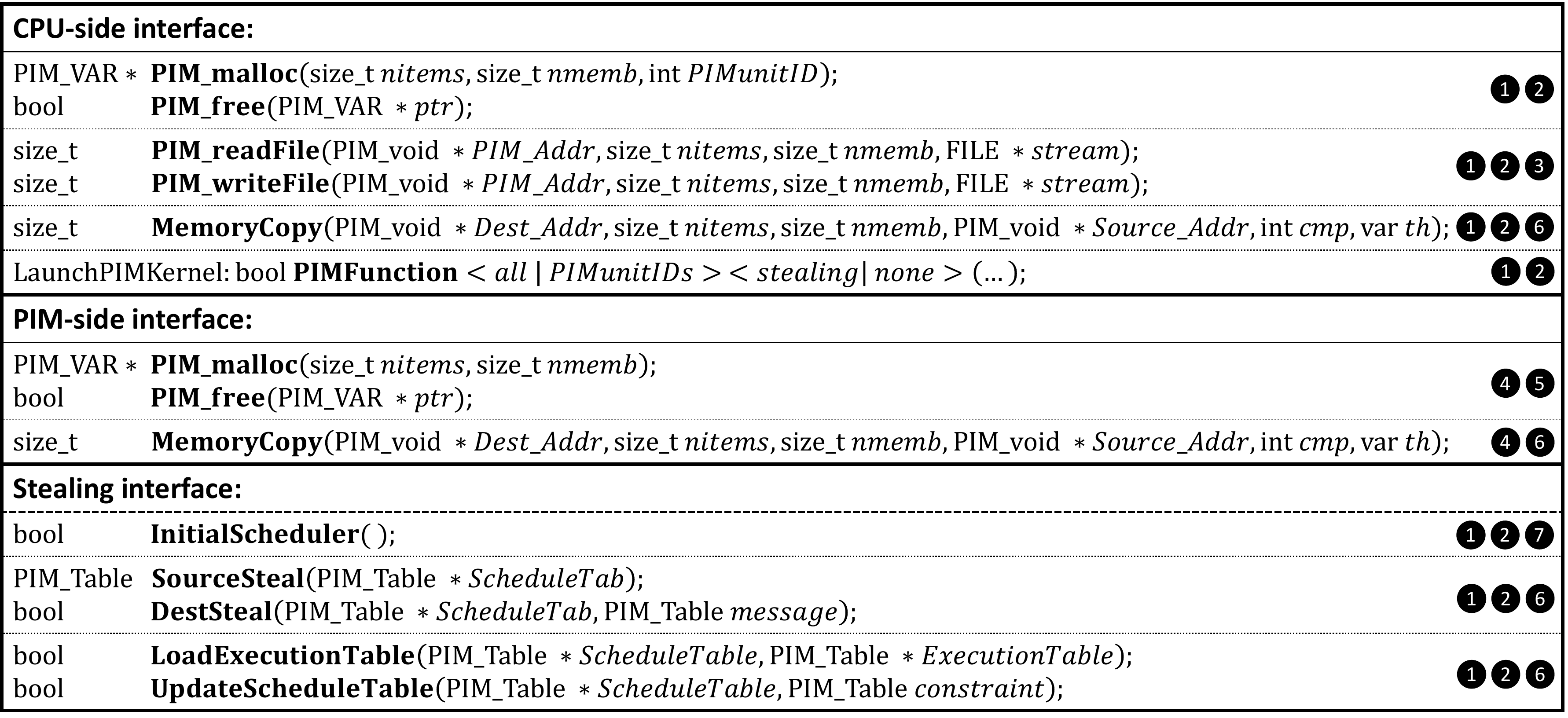}
\vspace{-0.5cm}
\caption{PIMMiner general programming interface.} 
\label{fig:interface}
\vspace{-0.5cm}
\end{figure}

\subsection{General Programming Interface}
\label{subsec:interface}

PIMMiner provides a set of programming interfaces that enable the use of new architecture. Additionally, we can utilize these basic interfaces to construct GPMI-PIM interfaces with the optimizations discussed above. The general interfaces are shown in Figure \ref{fig:interface}. 
The circled numbers indicate which connections in Figure \ref{fig:framework} (a) are used by this interface.

\subsubsection{CPU-side Interface}
CPU uses \emph{PIM\_malloc} and \emph{PIM\_free} interfaces to allocate and free space for the specified PIM unit. CPU could also directly manipulate the file from the disk into HBM-PIM memory with \emph{PIM\_readFile} and \emph{PIM\_writeFile}.
Additionally, CPU can initiate data copy on HBM-PIM from different bank groups with the \emph{MemoryCopy} interface. The \emph{MemoryCopy} interface also supports the access filter design in \S \ref{subsec:filter}: \emph{cmp} and \emph{th} can be set appropriately to filter out the unnecessary data in advance of execution.

When the CPU runs to \emph{LaunchPIMKernel}, it will wake up the PIM units to execute \emph{PIMFunction}. 
CPU allocates tasks with \emph{PIMFunction} to different PIM units. 
\emph{PIMFunction} has two options: the first option decides whether to load the code to all or specific PIM units; the second option decides whether stealing is enabled.

\subsubsection{PIM-side Interface}
PIM units can also use memory and data manipulation interfaces such as \emph{PIM\_malloc}, \emph{PIM\_free}, and \emph{MemoryCopy}. The difference is that one PIM unit can only directly malloc and free its own local memory. The \emph{MemoryCopy} supports one PIM unit read from another unit's data.

\subsubsection{Stealing Interface}
CPU side also has the following interfaces that support workload stealing. 
\emph{InitialScheduler} is used to initialize the stealing scheduler(Figure \ref{fig:framework}(c)) on each channel. Then, the CPU will load the \emph{SourceSteal}, \emph{DestSteal}, \emph{LoadExecutionTable} and \emph{UpdateScheduleTable} functions written by the programmer to each PIM unit as \emph{Steal Source Code}, \emph{Steal Dest Code}, \emph{Load Task Code} and \emph{Update Sche Tab Code}, respectively.
When executing \emph{PIMFunction}, the PIM unit will obtain the task ID according to the execution table. After finishing the current task, the PIM unit will update tables based on the workflow in Figure \ref{fig:Stealing workflow}.

\subsection{GPMI Interface}
\label{subsec:GPMI interface Implementation}

In this section, we describe our GPMI interface and the implementation details. 
There are two GPMI interfaces in PIMMiner, where \emph{PIMLoadGraph} loads the graph from the file to PIM memory automatically with data duplication optimization, and \emph{PIMPatternCount} counts the number of subgraphs matching the pattern \emph{p} in the graph \emph{G}.

\subsubsection{PIMLoadGraph \emph{interface}}
\emph{PIMLoadGraph}, as shown in Algorithm \ref{alg:Load Graph on PIM} is the preprocessing step operated by the CPU.
Conventionally, when loading a graph, the CPU will first read the whole graph into the main memory, then adjust the storage format of the graph, and finally copy the graph to another memory for execution. 
This procedure causes a high overhead between the main memory and other memory.
In contrast, with the \emph{PIMLoadGraph} interface, the CPU directly reads the graph to HBM-PIM memory from the disk without going through the main memory. 

Since the file reading process is sequential, when reading the neighbor list of each vertex to PIM, the CPU does not know all the data of the graph, so it cannot adjust the storage format of the graph in PIM memory. Therefore, we stipulate that the storage format of the graph in the file is CSR, which first stores the number of vertices, then stores the offset of each vertex in the \emph{ColIdx} array (\emph{RowPtr}), and finally stores their neighbor list (\emph{ColIdx}).

\begin{algorithm}[!h]
\caption{\emph{PIMLoadGraph} Interface (CPU)}
\label{alg:Load Graph on PIM}
\small
\begin{algorithmic}[]
\State{\textbf{Input:} \emph{Graph file F (CSR format, descending sorted degree)}} 
\State{\textbf{Output}: \emph{PIM Graph G}}
\end{algorithmic}
\begin{algorithmic}[1]
\State{Read \emph{RowPtr} array from \emph{F} to main memory}
\Comment{Load graph}
\For{each vertex $v$}
\State{Compute the length of $N(v)$ based on \emph{RowPtr}}
\State{Select PIM unit \emph{i} to store $N(v)$ (round-robin)}
\State{Allocate space in PIM unit \emph{i} using \emph{PIM\_malloc}}
\State{Read $N(v)$ from \emph{F} by \emph{PIM\_readFile}}
\EndFor 
\For{each PIM unit \emph{i}}
\Comment{Duplication}
\State{Compute $v_b$ using Algorithm \ref{alg:Critical Data}}
\For{each vertex $v<v_b$}
\State{Compute the length of $N(v)$ based on \emph{RowPtr}}
\State{Allocate space in PIM unit \emph{i} using \emph{PIM\_malloc}}
\State{Copy $N(v)$ from other PIM unit by \emph{MemoryCopy}}
\EndFor 
\EndFor 
\end{algorithmic}
\end{algorithm}

When loading the graph with \emph{PIMLoadGraph},
the CPU will first read the number of vertices and the \emph{RowPtr} array to the main memory (not HBM-PIM) (Line 1).
Then, it will calculate the length of the neighbor list of each vertex according to the \emph{RowPtr} array (Line 3), and choose a PIM unit to store the neighbor list (Line 4).
Here, we use the round-robin method to select the PIM unit.
The CPU will use the \emph{PIM\_malloc} interface to allocate space in the specified PIM unit (Line 5), and then use the \emph{PIM\_readFile} interface to load the neighbor list $N(v)$ from the file into the specified PIM unit. (Line 6)


\noindent \textit{Selective Vertex Duplication Optimization: }
When the CPU finishes loading graphs, it will use free PIM memory space to selectively duplicate vertices, which could further enhance the utilization of the local bank groups. 
We use vertex degree as the indicator of access frequency, and duplicate high-degree vertices in the free memory as much as possible. 
We assume that the vertices have been sorted in descending order.
The CPU then computes the vertices that can be duplicated in each PIM unit through Algorithm \ref{alg:Critical Data} (Line 8), and use the CPU-side interfaces \emph{PIM\_malloc} and 
\emph{MemoryCopy} to copy these neighbor lists to the corresponding PIM unit (Lines 10-12).



\begin{algorithm}[!ht]
\caption{Calculate Duplicated Data in the PIM Unit}
\label{alg:Critical Data}
\small
\begin{algorithmic}[]
\State{\textbf{Input:} \emph{Sorted graph G}, \emph{Remaining memory size in PIM unit M}} 
\State{\textbf{Output}: \emph{Boundary vertex $v_b$}}
\end{algorithmic}
\begin{algorithmic}[1]
\State{\emph{UsedMem} $\gets$ 0}
\For{\emph{i} = 0 \textbf{to} \emph{G.v} - 1}
\If{(\emph{UsedMem} + \emph{G.$v_i$.neighborSize}) $\leq$ \emph{M} }
\State{\emph{UsedMem} $\gets$ \emph{UsedMem} + \emph{G.$v_i$.neighborSize}}
\Else{}
\State{break}
\EndIf
\EndFor 
\State{$v_b$ $\gets$ \emph{i}}
\end{algorithmic}
\end{algorithm}

\subsubsection{PIMPatternCount \emph{interface}}

Since pattern counting in GPMI has a serious load imbalance problem on PIM (\S \ref{subsec:load imbalance}), the \emph{PIMPatternCount} interface first needs to set the stealing parameters and then initiates the PIM units execution with \emph{PIMFunction}.

We have described the stealing interface and implementation details in Figure \ref{fig:interface} and section \ref{subsubsec:Implementation Details to Support GPMI}.

The implementation of \emph{PIMFunction} depends on the pattern. The PIM unit will follow the steps below to finish counting a certain pattern: 
\begin{itemize}
  \item Use the execution table $T_{exe}$ to obtain the vertex $v_i$.
  \item Load the neighbor lists in the \emph{i}th for loop, if $T_{exe}[i+1]=0$.
  \item When accessing $N(V)$, if $v<v_b$, $N(v)$ has a copy in the local memory; otherwise, call the \emph{MemoryCopy} interface to read from other units.
  \item If $v_i$ has restrictions such as $v_i<c$, then when calling the \emph{MemoryCopy} interface, the PIM unit needs to set \emph{cmp} to $<$ and \emph{th} to $c$, so that the filter can filter out unnecessary data in advance.
\end{itemize}

\section{Experimental Methodologies}
\label{subsec:setup}

\noindent \textbf{Graph datasets.}
We use the graph datasets in Table~\ref{tab:Graph Datasets} for evaluation.
These datasets are used for most GPMI systems, such as Arabesque~\cite{zaki2015arabesque}, RStream~\cite{wang2018rstream}, AutoMine~\cite{mawhirter2019automine},
Gramer~\cite{yao2020locality},
GraphPi~\cite{shi2020graphpi} and
FlexMiner~\cite{chen2021flexminer}.
Before execution, we sort the vertices based on their degree from largest to smallest (the id of the vertex with the highest degree is 0).

\begin{table}[t]
	\centering  
	\small
	\caption{Graph Datasets}  
	\label{tab:Graph Datasets}  
	\resizebox{0.9\linewidth}{!}{
	\begin{tabular}{c|c|c|c|c}  
	\hline
	    \textbf{Graph} & |$V$| & |$E$| & Size & Max Degree\\
		\hline  
	    CiteSeer(CI)~\cite{elseidy2014grami} & 3,264 & 4,536 & 84KB & 99  \\ 
	    P2P(PP)~\cite{leskovec2014snap} & 10.9K & 40.0K & 620K & 103\\
	    Astro(AS)~\cite{leskovec2014snap} & 18.8K & 198K & 5.3M & 504 \\
		MiCo(MI)~\cite{elseidy2014grami} & 100K & 1.08M & 18MB & 1359  \\
		com-Youtube(YT)~\cite{leskovec2014snap} & 1.13M & 2.99M & 57MB & 28,754  \\
		cit-Patents(PA)~\cite{leskovec2014snap} & 3.77M & 16.52M & 332MB & 793 \\
		soc-LiveJournal1(LJ)\cite{leskovec2014snap} & 4.85M & 43.11M & 1.2G & 20,334  \\
		\hline
	\end{tabular}
	}
\end{table}

\begin{table}[t]
	\centering  
	\caption{Simulated system configuration.}  
	\label{tab:Simulated Configuration}  
	\resizebox{\linewidth}{!}{
	\begin{tabular}{|c|l|}  
	    \hline
	    \multicolumn{2}{|c|}{\textbf{PIM Simulator}} \\ \hline
	    \multirow{2}{*}{PIM Execution Units} & 128 in-order cores, 250MHz, 4-issue, \\
     & 32 FPUs each core, total 1024 GFLOPs~\cite{lee2021hardware}\\ \hline
		L1I Cache & private, 32KB, 4-way, 4-cycle, 64B, 16 MSHRs~\cite{gao2015practical}  \\ \hline
		L1D Cache & private, 32KB, 8-way, 4-cycle, 64B, 16 MSHRs~\cite{gao2015practical}  \\ \hline
		Memory & 1GHz 3D Memory Stack~\cite{lee2021hardware}\\ \hline
		\hline  
		\multicolumn{2}{|c|}{\textbf{3D Memory Stack}} \\ \hline
		\multirow{2}{*}{Organization} & 4GB, 4 layers, 32 channels, 8 banks per channel,\\
		& 4 PIM Units per channel \\ \hline
		\multirow{2}{*}{Timing Parameters} & $t_{CK} = 1$ ns, $t_{RAS} = 27$ ns, $t_{RCD} = 14$ ns, \\ & $t_{CL} = 14$ ns, $t_{WR} = 15$ ns, $t_{RP} = 14$ ns~\cite{nai2017graphpim}\\ \hline
		\multirow{2}{*}{External links} & 4 half-duplex serialized links, 8 lanes/link, 15 Gb/s lane speed, \\ 
		& total 120 GB/s bandwidth, 500-cycle latency  \\ \hline
		\multirow{2}{*}{Internal links} & 32 links, 8 Bytes/cycle, 8 GB/s per link, total 256 GB/s bandwidth,\\ 
		&  40-cycles bank latency, 140 channel latency\\
		\hline
		\multirow{2}{*}{On-chip links} & 128 links, 8 Bytes/cycle, 8 GB/s per link, total 1 TB/s bandwidth, \\
		& 10-cycle in-bank latency\\
		\hline
	\end{tabular}
	}
	\vspace{-0.3cm}
\end{table}

\noindent \textbf{GPMI applications.}
We evaluate the following GPMI applications: 3-size motif counting (3-MC), 3 to 5-size clique counting (3-CC, 4-CC, 5-CC), 4-size diamond (4-DI), and cycle(4-CL) patterns. These patterns are described in Figure~\ref{fig:pattern} and are commonly used in other GPMI systems~\cite{shi2020graphpi,chen2021flexminer,chen2022fingers}.

\begin{figure*}[ht]
\centering 
\includegraphics[width=\linewidth]{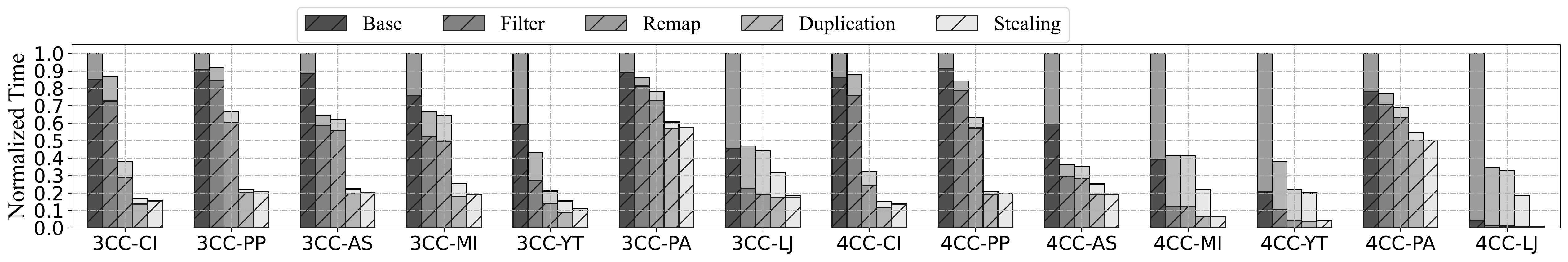}
\includegraphics[width=\linewidth]{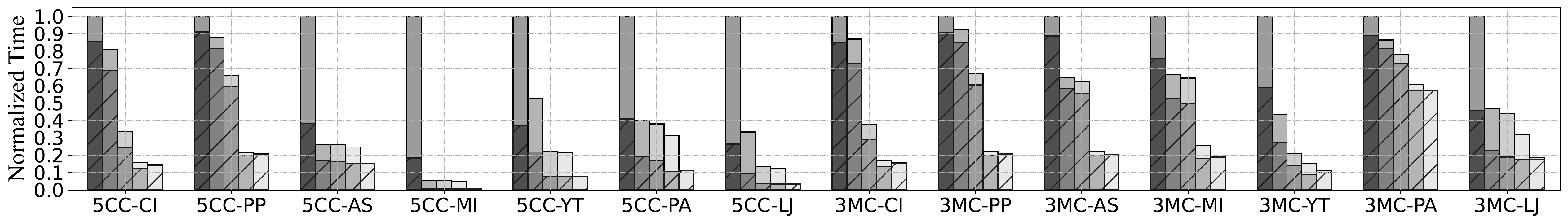} 
\includegraphics[width=\linewidth]{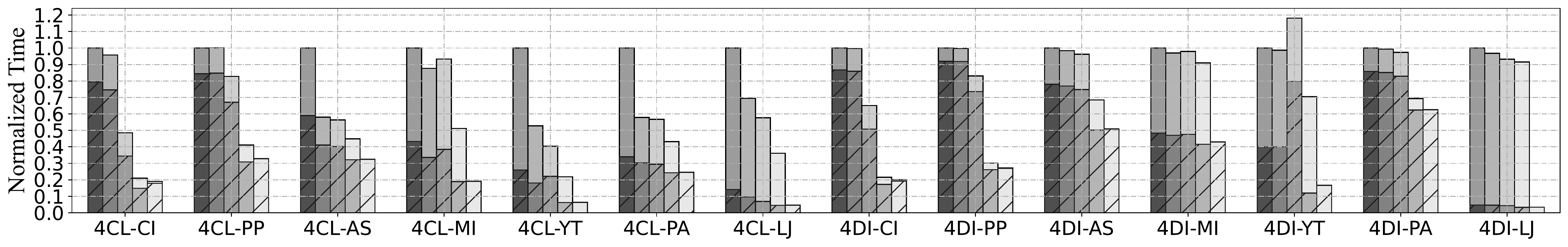} 
\caption{Performance of PIMMiner with the effectiveness of proposed optimizations. In each bar, we show the average time across cores (the solid line) and the total execution time(top of bar).} 
\label{fig:performance}
 \vspace{-0.2in}
\end{figure*}

\noindent \textbf{Compared GPMI software systems.}
\textit{GraphPi}~\cite{shi2020graphpi} is a graph mining system on CPUs. 
GraphPi is aimed at complex patterns, it determines the order of each vertex of the pattern according to the characteristics of the graph and the pattern.
We run GraphPi with the source code~\cite{shi2020graphpiproject} using 96 threads.
\textit{AutoMine}~\cite{mawhirter2019automine} is a widely-used pattern-enumeration method, and our GPMI algorithm is based on this method. We requested the code of Automine from the author. When running the original Automine code, we found that Automine uses multiple function calls for the generality, which greatly reduces the performance, and the load is
extremely imbalanced when multi-threaded. So, we rewrote the code using the order of each vertex of the pattern in GraphPi \protect\footnotemark[2], and use it for PIM acceleration. 
We show the runtime of original(ORG) and optimized(OPT) Automine in the evaluation section.
The results in Table \ref{tab:Performance 128-core CPU and 128-core PIM 4-CC} are also run with optimized(OPT) AutoMine.

\footnotetext[2]{https://anonymous.4open.science/r/Optimized-AutoMine/README.md}

\noindent \textbf{Compared GPMI acceleration architectures.}
\textit{DIMMining} \cite{dai2022dimmining} and \textit{NDMiner} \cite{talati2022ndminer} are two recent PIM/NDP architectures designed for accelerating GPMI applications. 
The results show that the performance of \text{DIMMining} and \text{NDMiner} is higher than the previous accelerators\cite{yao2020locality,chen2021flexminer}, so we aim to use these two designs as the baselines of GPMI accelerators.
For \text{DIMMining}, we use the numerical values reported in \cite{dai2022dimmining}, and we also requested the execution time of \text{NDMiner} from its authors.
To ensure the fairness of the experimental results, we scaled their results to the peak performance of 1024 Gflops, which is the same as our \text{PIMMiner}.

\noindent \textbf{System and architecture configurations.}
For AutoMine and GraphPi, we run the experiments on a real system with Intel(R) Xeon(R) Gold 6240R 2.4GHz Processor (total 48 physical cores, 96 hyperthreads) with 36MB LLC and 192GB 3.2GHz DDR4 memory.
According to Intel's official documentation, the peak of this processor is 1152 Gflops, which is very close to the 1024 Glops of PIMMiner architecture.
For PIMMiner, we use ZSim~\cite{sanchez2013zsim} with Ramulator~\cite{kim2015ramulator} to simulate HBM-PIM systems. 
We modify Zsim to generate traces for PIM when executing the \textit{nest\_for\_loop} function.  
We also modify Ramulator to support 128 PIM cores (following Samsung HBM-PIM parameters\cite{kwon202125,lee2021hardware}) with L1i and L1d caches. All caches use the LRU policy.
The memory access latency in 3D-stack DRAM is obtained from \cite{hadidi2018performance}.
Table~\ref{tab:Simulated Configuration} shows the detailed configurations of our simulated system.
We set the overhead of workload stealing to 2 times of remote memory access latency ($2 \times 140 = 280$ memory cycles).
Due to the long execution time of the cycle-accurate simulator, for larger graphs (MI, YT, PA, and LJ), we sample the vertices $v_0$ in the first level during trace generation. The sample ratio is set so that each simulation can finish within ten hours. Other works using cycle-accurate simulation also adopt the similar sampling methodology \cite{besta2021sisa}.

\section{Evaluation Results}
\subsection{Overall Performance}
\label{subsec: overall performance}

In this section, we first report the overall performance of PIMMiner with proposed optimizations. We then compare the performance PIMMiner with other GPMI systems.
To match the names in the tables and figures, in this section, we will use abbreviations to denote optimizations.
Among them,
\emph{Filter} refers to the application-aware memory access filter in Section \ref{subsec:filter}, \emph{Remap} refers to PIM-friendly data mapping in Section \ref{subsec:HBM-PIM memory controller}, \emph{Duplication} refers to the vertex duplication optimization when calling the \emph{PIMLoadGraph} interface in Section \ref{subsec:GPMI interface Implementation}, and finally \emph{Stealing} refers to the workload stealing scheduler described in Section \ref{subsec:steal}.

\subsubsection{PIMMiner performance with proposed optimizations}

Figure \ref{fig:performance} shows the performance improvement of PIMMiner by enabling the optimizations proposed in Section \ref{sec:PIMMiner Design} one by one.
In Figure \ref{fig:performance}, the top of the bar represents the overall execution time, the solid line represents the average execution time for all PIM units.
The difference between the two indicates how imbalanced the workload is.

As shown in Figure \ref{fig:performance}, each of our optimizations reduces the execution time by a fraction; both filter and stealing designs can help balance workloads across cores.
The filter design achieves 2.01x average speedup and 17.57x maximum speedup over our baseline PIM. 
Remap achieves 1.38x average speedup and 2.74x maximum speedup over the filter.
Duplication achieves 1.84x average speedup and 3.05x maximum speedup over the remap.
And the stealing achieves 3.01x average speedup and 26.87x maximum speedup over the duplication optimization.
Overall, by enabling all optimizations, PIMMiner achieves 12.74x average speedup and 113.76x maximum speedup over the baseline PIM implementation.

In Figure \ref{fig:performance}, the execution time of 4CL-MI and 4DI-YT increases with remapping.
The reason is that remapping increases the possibility of different PIM units accessing the same bank at the same time (for example, all units access $N(0$) on bank $0$ at the same time), resulting in memory access congestion on a few banks.
Such an issue can be fixed with duplication optimization so that PIM units access their own bank in most cases, and the possibility of PIM units accessing the same memory bank at the same time can be reduced.
PIMMiner with all proposed optimizations can cohesively work together to address the memory access distribution and load imbalance issues discussed in Section \ref{sec:motivation}.

\subsubsection{Comparison with other systems}

We compare our PIMMiner with other GPMI systems, and the results are shown in Table \ref{tab:Graph Mining Systems Comparison}.
We report the results of DIMMining on \emph{PP}, \emph{AS}, and \emph{MI}, and NDMiner results on the \emph{PA} graph, since these graphs are the same as the ones we use (Table \ref{tab:Graph Datasets}).
In Table \ref{tab:Graph Mining Systems Comparison}, the shortest execution time for each task is in bold, and we can observe that our PIMMiner is the fastest in most cases.
If not the fastest, our PIMMiner is also in the same order of magnitude as the fastest execution time.



First, we compare our PIMMiner with the software GPMI systems.
Among the three software GPMI systems, our optimized Automine is always the fastest one except for a few cases.
Therefore, we choose the optimized AutoMine as the base algorithm of our PIMMiner.
For GraphPi, our PIMMiner achieves 549.41x average speedup and 8755x maximum speedup.
For the original AutoMine, our PIMMiner achieves 710.17x average speedup and 2826x maximum speedup.
And for the optimized AutoMine, our PIMMiner achieves 132.19x average speedup and 1312x maximum speedup.


Then, we will compare our PIMMiner with the PIM or NDP-based GPMI accelerators.
Our PIMMiner achieves 2.70x average speedup and 6.33x maximum speedup over DIMMining, 59.30x average speedup and 154.87x maximum speedup over NDMiner.
It should be noted that PIMMiner uses much more general processing units and interfaces, while both processing units of DIMMining and NDMiner are specifically optimized for GPMI applications.
PIMMiner has achieved high performance even without optimizing the processor unit. We believe that PIMMiner can be further optimized with set-centric computing units like the ones in SISA\cite{besta2021sisa}, FlexMiner\cite{chen2021flexminer}, DIMMining\cite{dai2022dimmining} and NDMiner\cite{talati2022ndminer}. We leave this direction as our future work.



\begin{table}[!h]
	\centering  
	\small
	\caption{Graph Mining Systems Comparison (in seconds)\protect\footnotemark[3]}  
	\label{tab:Graph Mining Systems Comparison}  
	\resizebox{\linewidth}{!}{
	\begin{tabular}{c|c|c|c|c|c|c}  
	\hline
	    Pattern & G & GraphPi & AM(ORG) & AM(OPT) & DIM\&ND & PIMMiner\\
		\hline  
		\multirow{7}{*}{3-CC} & CI & 4.64E-02 & 1.45E-02 & 4.87E-03 & -- & \textbf{5.30E-06}\\
		& PP & 6.72E-02 & 3.57E-02 & 9.54E-03 & 3.82E-05 & \textbf{3.36E-05} \\
		& AS & 7.43E-02 & 3.22E-01 & 1.12E-02 & 6.14E-04 & \textbf{2.22E-04}\\
		& MI & 9.93E-02 & 2.53 & 2.69E-02 & 3.77E-03 & \textbf{1.46E-03}\\
		& YT & 2.32E-01 & 23.39 & 1.34E-01 & -- & \textbf{1.21E-02}\\
		& PA & 2.32E-01 & 21.84 & 1.98E-01 & 3.68E-01 & \textbf{3.35E-02}\\
		& LJ & 2.32 & 186.61 & 1.24 & -- & \textbf{1.59E-01}\\
		\hline
		\multirow{7}{*}{4-CC} & CI & 1.49E-02 & 1.07E-03 & 4.36E-04 & -- & \textbf{5.86E-06}\\
		& PP & 1.23E-02 & 1.00E-02 & 3.79E-03 & 4.10E-05 & \textbf{3.38E-05}\\
		& AS & 1.91E-02 & 6.29E-01 & 8.06E-02 & 3.79E-03 & \textbf{7.86E-04}\\
		& MI & 2.37E-01 & 11.82 & 2.39E-01 & 5.33E-02 & \textbf{2.77E-02}\\
		& YT & 2.01E-01 & 3.05 & 2.08E-01 & -- & \textbf{7.48E-02}\\
		& PA & 2.94E-01 & 3.47 & 2.40E-01 & 7.38E-01 & \textbf{3.47E-02}\\
		& LJ & 6.53 & 256.42 & 2.78 & -- & \textbf{1.16}\\
		\hline
		\multirow{7}{*}{5-CC} & CI & 1.62E-02 & 2.08E-03 & 4.70E-04 & -- & \textbf{6.02E-06}\\
		& PP & 1.22E-02 & 8.81E-03 & 3.79E-03 & 4.13E-05 & \textbf{3.39E-05}\\
		& AS & 6.10E-02 & 6.31 & 1.60E-01 & 2.42E-02 & \textbf{4.68E-03} \\
		& MI & 10.36 & 2110.88 & 4.35 & 1.86  & \textbf{7.47E-01}\\
		& YT & 4.53E-01 & 97.94 & 3.12E-01& -- & \textbf{2.24E-01}\\
		& PA & 1.61E-01 & 5.17 &  1.90E-01 & 1.47 & \textbf{1.62E-02}\\
		& LJ & 210.01 & 5.15E+04 & 99.64 & -- & \textbf{95.10}\\
		\hline
		\multirow{7}{*}{3-MC} & CI & 1.84E-02 & 1.65E-02 & 1.43E-02 & -- & \textbf{1.09E-05} \\
		& PP & 2.12E-02 & 4.56E-02 & 1.70E-02 & 1.14E-04 & \textbf{4.96E-05}\\
		& AS & 3.32E-02 & 4.08E-01 & 1.76E-02 & 2.18E-03 & \textbf{3.44E-04}\\
		& MI & 3.69E-02 & 3.23 & 4.26E-02 & 1.48E-02 & \textbf{3.07E-03}\\
		& YT & 2.32E-01 & 25.39 & 4.48E-01 & -- & \textbf{1.75E-01}\\
		& PA & 2.76E-01 & 27.07 & 3.28E-01 & -- & \textbf{4.34E-02} \\
		& LJ & 1.04 & 218.09 & 1.72 & -- & \textbf{3.56E-01}\\
        \hline
		\multirow{7}{*}{4-DI} & CI & 1.03E-02 & 2.43E-03 & 9.39E-03 & -- & \textbf{7.21E-06} \\
		& PP & 1.18E-02 & 1.13E-02 & 9.83E-03 & 9.55E-05 & \textbf{4.64E-05}\\
		& AS & 1.70E-02 & 1.04 & 1.02E-02 & 1.49E-03 & \textbf{1.22E-03}\\
		& MI & 7.28E-02 & 25.49 & 2.34E-01 & \textbf{1.18E-02} & 3.01E-02\\
		& YT & 9.25E-02 & 8.78 & 1.23E-01 & -- & \textbf{8.30E-02}\\
		& PA & 1.63E-01 & 11.7 & 1.37E-01 & 8.08E-01 & \textbf{4.34E-02} \\
		& LJ & 1.9 & 705.4 & 5.54 & -- & \textbf{1.02}\\
        \hline
		\multirow{7}{*}{4-CL} & CI & 1.09E-02 & 2.52E-03 & 1.50E-03 & -- & \textbf{6.54E-06} \\
		& PP & 1.23E-02 & 2.78E-02 & 1.03E-02 & -- & \textbf{6.60E-05}\\
		& AS & 3.26E-02 & 3.17E-01 & 3.26E-02 & -- & \textbf{2.99E-03}\\
		& MI & 4.31E-01 & 3.21 & 2.18E-01 & -- & \textbf{9.19E-02}\\
		& YT & 2.29 & 18.83 & 2.54  & -- & \textbf{2.80E-01}\\
		& PA & 4.13E-01 & 28.75 & 7.67E-01 & 9.664 & \textbf{6.24E-02} \\
		& LJ & 31.09 & 417.03 & 40.09 & -- & \textbf{6.01}\\
        \hline
	\end{tabular}
 	}
\end{table}

\subsection{In-depth Study}
\label{subsec: Benefit Of Our Design}

In this section, we use 4-CC as an example to compare the performance improvement and the remote memory access reduction brought by each optimization. 

\subsubsection{Benefits of the Application-aware Memory Filter}

First, we discuss the impact of the filter on memory access.
Table \ref{tab:Benefit Of Conditional Access Filtering} shows the memory access size of 4-CC before and after filtering.
As shown in Table \ref{tab:Benefit Of Conditional Access Filtering}, the greater the ratio of access memory reduction, the higher the performance improvement brought by filtering.
In addition, we find that the size of the memory access is much larger than the size of the graph itself. For example, \emph{AS} is a small graph that has a total memory access size (166MB) of about 31x over the graph size (5.3MB).
Finally, according to Table \ref{tab:Benefit Of Conditional Access Filtering}, the larger the total memory access is, the more data is filtered out.
For the graphs with more than 100 MB of memory footprint, the filter design reduces memory access by more than 60\% and execution time by more than half.
We believe this is due to the slight cache pollution on the small graphs, resulting in fewer memory accesses that need to be filtered.


\begin{table}[!h]
	\centering  
	\small
	\caption{Benefit of the filter in 4-CC (TM is total memory access size; RM is the filtered memory access size; Ratio is reduced memory size ratio). }  
	\label{tab:Benefit Of Conditional Access Filtering}  
	\resizebox{\linewidth}{!}{
	\begin{tabular}{c|c|c|c|c|c|c|c}  
	\hline
	     & CI & PP & AS & MI & YT & PA & LJ \\
		\hline  
        TM & 1.3MB & 8.2MB & 166MB & 2.1GB & 1.2GB & 48MB & 707MB \\
        FM & 1.0MB & 5.5MB & 36.9MB & 316MB & 474MB & 30MB & 144MB \\ 
        Ratio & 22\% & 33\% & 78\% & 85\% & 59\% & 38\% & 80\% \\ 
        \textit{Speedup} & 1.13x & 1.19x & 2.76x & 2.41x & 2.64x & 1.30x & 2.90x \\
		\hline
	\end{tabular}
	}
 \vspace{-0.1in}
\end{table}

\footnotetext[3]{AM(ORG) = original Automine; AM(OPT) = optimized Automine. DIM\&ND reports the results of graphs PP, AS, MI from DIMMining, and graph PA from NDMiner. MI uses 10\% sampling, YT and PA use 1\% sampling, and LJ uses 0.1\% sampling.}

\subsubsection{Benefit of PIM-friendly Data Mapping}

Next, we analyze the benefit of remapping. We measure the memory accesses that fall into the local banks and compute the local access ratio in Table \ref{tab:Benefit Of Rempa and Dup}.
With the new mapping scheme, the ratio of local memory accesses increases significantly, especially for graphs \emph{CI} and \emph{YT}, which shows that those two graphs can achieve good locality of memory access without duplication.
For other graphs, although remapping has only a little improvement in local access ratio and performance, PIM-friendly data mapping is the basis for vertex duplication optimization.

\begin{table}[!h]
	\centering  
	\small
	\caption{Local access ratio and speedup improvement with remapping and duplication in 4-CC (Baseline has applied the filter optimization).}  
	\label{tab:Benefit Of Rempa and Dup}  
	\resizebox{\linewidth}{!}{
	\begin{tabular}{c|c|c|c|c|c|c|c}  
	\hline
	     & CI & PP & AS & MI & YT & PA & LJ \\
        \hline
        Baseline & 1.36\% & 1.36\% & 1.78\% & 2.03\% & 1.22\% & 1.33\% & 5.74\% \\
        \hdashline[1pt/1pt]
        Remap & 86.86\% & 60.19\% & 32.68\% & 19.31\% & 98.62\% & 50.34\% & 69.23\% \\
        \textit{Speedup} & 2.74x & 1.33x & 1.03x & 1.01x & 1.73x & 1.12x & 1.05x \\
        \hdashline[1pt/1pt]
        Duplication & 100\% & 100\% & 100\% & 100\% & 100\% & 66.27\% & 90.51\% \\
        \textit{Speedup} & 2.12x & 3.04x & 1.39x & 1.86x & 1.09x & 1.26x & 1.75x \\
        \hline
	\end{tabular}
	}
\end{table}


\subsubsection{Benefit of Vertex Duplication Optimization}
Then, we show the improvement of data locality through duplication optimization.
First, according to Algorithm \ref{alg:Critical Data}, we calculate the maximum number of the neighbor list duplication that each bank group can store.
The first five graphs can duplicate the entire graph in each bank group since they are very small.
The last two graphs \emph{PA} and \emph{LJ} are large; therefore, the algorithm selects the top 5\% and 0.25\% neighbor list for duplication.
With vertex duplication, the local access memory access ratio of the first 5 graphs reaches 100\%, \emph{LiveJournal} also reaches 90\%, and \emph{Patents} reaches 66\% in Table \ref{tab:Benefit Of Rempa and Dup}.
The speedup achieved through duplication is high, even for the larger graphs that cannot fully duplicate the neighbor lists, such as \emph{LiveJournal} and \emph{Patents}.

\subsubsection{Benefit of Workload Stealing Scheduler}
In Figure \ref{fig:performance}, as the pattern size increases, the gap between the execution time and average time becomes larger, which means that the workload becomes more and more imbalanced.
The filter reduces load imbalance issues for several applications; for larger patterns, our workload stealing can further mitigate the load imbalance issues dynamically.
We analyze 4-CC in Table \ref{tab:Benefit Of PIM-aware Work Stealing}.
The first row shows the ratio of the execution time to the average time after the first three optimizations.
We can observe that even with the prior optimizations, \emph{MI}, \emph{YT}, and \emph{LJ} still have significant load imbalance issues.
With the workload stealing scheduler, the difference between execution time and
average time becomes negligible.
Our stealing on 4-CC can achieve up to 20x speedup.

\begin{table}[!h]
	\centering  
	\small
	\caption{Benefit of workload stealing in 4-CC.}  
	\label{tab:Benefit Of PIM-aware Work Stealing}  
	\resizebox{\linewidth}{!}{
	\begin{tabular}{c|c|c|c|c|c|c|c}  
	\hline
	     & CI & PP & AS & MI & YT & PA & LJ \\
		\hline  
		Exe/Avg (no steal) & 1.28 & 1.09 & 1.33 & 3.46 & 5.24 & 1.09 & 22.23 \\
		Exe/Avg (steal) & 1.06 & 1.004 & 1.001 & 1.001 & 1.01 & 1.001 & 1.003 \\
		\textit{Speedup} & 1.07x & 1.05x & 1.30x & 3.38x & 4.92x & 1.08x & 20.45x \\
		
		\hline
	\end{tabular}
	}
 \vspace{-0.1in}
\end{table}

\section{Related Work}
\subsection{Software GPMI systems}
There are many software GPMI systems emerged in recent years.  Arabesque\cite{zaki2015arabesque}, RStream \cite{wang2018rstream}, Pangolin\cite{chen2020pangolin}, DistGraph~\cite{talukder2016distributed}, ScaleMine~\cite{abdelhamid2016scalemine}, Fractal\cite{dias2019fractal} are all exploration based, which are much less efficient compared to Automine \cite{mawhirter2019automine}, which enumerates all the unlabeled patterns of a particular size and match them one-by-one on a graph. GraphZero\cite{mawhirter2019graphzero} and GraphPi\cite{shi2020graphpi} are also based on the mining algorithm proposed in Automine with more compiler support and computation redundancy elimination.
We choose Automine, one of the most efficient software-based pattern-enumeration systems as PIMMiner's baseline GPMI algorithm. PIMMiner can substantially outperform all software-based solutions by efficiently using PIM architecture. We believe that PIMMiner can still gain significant performance benefits with new GPMI algorithms, since our proposed optimizations are very general and efficient.

\subsection{Hardware acceleration for GPMI}
Gramer\cite{yao2020locality} was one of the first graph mining accelerators that use a specialized cache design to improve the locality of GPMI applications. Gramer is based on the exhaustive-check method, therefore, it cannot outperform software pattern-enumeration methods like Automine\cite{mawhirter2019automine}.
FlexMiner\cite{chen2021flexminer} and Fingers\cite{chen2022fingers} are two recent accelerator designs for GPMI applications, which utilize multiple PEs and explore the internal parallelism in GPMI applications.
\textit{IntersectX}~\cite{rao2020intersectx} is an accelerator for GPMI applications with streaming instruction set extension and architectural supports based on a conventional processor.
\textit{Mint}~\cite{talati2022mint} is a novel accelerator to mine the temporal motifs efficiently, which is similar to GPMI, but with stricter constraints on edges.

SISA~\cite{besta2021sisa} utilizes PIM architecture with specialized set-operation ISAs and hardware units to accelerate set operations. We are not able to compare with \textit{SISA}~\cite{besta2021sisa}, because it uses different applications and does not compare with existing GPMI systems or report relative speedup. 
Su et.al \cite{su2021exploring} propose to use general PIM to accelerate GPMI applications, however, the issue of load imbalance is not fully addressed in the work.
NDMiner\cite{talati2022ndminer} and DIMMining\cite{dai2022dimmining} are the two recent works using the in-memory specialized accelerator to accelerate GPMI applications; our PIMMiner uses much simpler PIM hardware while achieving higher performance gain.

\section{Conclusions}
We propose PIMMiner, a new graph mining framework that can leverage PIM architecture for high-performance computing. 
We identify that PIM has the potential to accelerate graph mining applications if the graph data and workload scheduling are done by considering the internal architectural properties of HBM-PIM.
With the PIMMiner framework, we can achieve optimized data placement and execution flow with much more general processing units and interfaces. PIMMiner gains 132x-710x average speedup over existing software GPMI systems and improves PIM baseline performance by 15.91x.
We believe that PIMMiner can be further optimized with set-centric computing units like the ones in SISA\cite{besta2021sisa}, FlexMiner\cite{chen2021flexminer}, DIMMining\cite{dai2022dimmining} and NDMiner\cite{talati2022ndminer}.

\bibliographystyle{IEEEtranS}
\bibliography{ref}


\end{document}